\documentclass{article}

\usepackage{arxiv}

\usepackage[utf8]{inputenc} 
\usepackage[T1]{fontenc}    
\usepackage{hyperref}       
\usepackage{url}            
\usepackage{booktabs}       
\usepackage{amsfonts}       
\usepackage{nicefrac}       
\usepackage{microtype}      
\usepackage{lipsum}		
\usepackage{graphicx}
\usepackage{natbib}
\usepackage{doi}

\title{CLOINet: Ocean state reconstructions through
remote-sensing, in-situ sparse observations
and Deep Learning}

\author{ 
    \hspace{1mm}Eugenio Cutolo \\
	IMEDEA (CSIC-UIB), \\
    Esporles, Spain, \\
    \texttt{e.cutolo@imedea.uib-csic.es} \\
	\And
    \hspace{1mm}Ananda Pascual \\
	IMEDEA (CSIC-UIB), \\
    Esporles, Spain, \\
    \texttt{ananda.pascual@imedea.uib-csic.es} \\
	\And
    \hspace{1mm}Simon Ruiz \\
	IMEDEA (CSIC-UIB), \\
    Esporles, Spain, \\
    \texttt{simon.ruiz@imedea.uib-csic.es } \\
	\And
    \hspace{1mm}Nikolaos Zarokanellos \\
    SOCIB,\\
    Palma, Spain\\
    \texttt{nzarokanellos@socib.es} \\
	\And
    \hspace{1mm}Ronan Fablet \\
	IMT Atlantique, \\
    CNRS UMR Lab-STICC, \\
    Brest, France \\
	\texttt{ronan.fablet@imt-atlantique.fr} \\
}

\renewcommand{\shorttitle}{CLOINet: Ocean State Reconstructions}

\hypersetup{
pdftitle={CLOINet: Ocean State Reconstructions},
pdfauthor={Eugenio Cutolo},
pdfkeywords={deep-learning, ocean, remote-sensing, SST, SSH, gliders, OSSE},
}

\begin{document}
\maketitle

\begin{abstract}
Combining remote-sensing data with in-situ observations to achieve a comprehensive 3D reconstruction of the ocean state presents significant challenges for traditional interpolation techniques. To address this, we developed the CLuster Optimal Interpolation Neural Network (CLOINet), which combines the robust mathematical framework of the Optimal Interpolation (OI) scheme with a self-supervised clustering approach. CLOINet efficiently segments remote sensing images into clusters to reveal non-local correlations, thereby enhancing fine-scale oceanic reconstructions. We trained our network using outputs from an Ocean General Circulation Model (OGCM), which also facilitated various testing scenarios. Our Observing System Simulation Experiments aimed to reconstruct deep salinity fields using Sea Surface Temperature (SST) or Sea Surface Height (SSH), alongside sparse in-situ salinity observations. The results showcased a significant reduction in reconstruction error up to $40\%$ and the ability to resolve scales $50\%$ smaller compared to baseline OI techniques. Remarkably, even though CLOINet was trained exclusively on simulated data, it accurately reconstructed an unseen SST field using only glider temperature observations and satellite chlorophyll concentration data. This demonstrates how deep learning networks like CLOINet can potentially lead the integration of modeling and observational efforts in developing an ocean digital twin. 
\end{abstract}


\section{Introduction}
Nowadays, there is an increased consciousness of the role played by the ocean in many crucial aspects of human safety, health, and well-being due to the cumulative impacts of climate change, unsustainable exploitation of marine resources, pollution, and uncoordinated development (UNESCO, 2019,\cite{Pascual2021White2030}). In response to these challenges, which UNESCO has encapsulated in 10 objectives for the Ocean Decade (2021-2030), the European Union is endeavoring to develop a digital twin of the ocean. The concept of digital twins involves creating a digital representation of real-world entities or processes, based on both real-time and historical observations, to depict the past and present and to model potential future scenarios.

In the ocean case and especially to address climate change-related concerns, one major challenge is understanding the state and evolution of the ocean's interior. Its stratification significantly influences large-scale integrated variables like ocean heat content, acidification, and oxygenation \citep{Wang2018ConsensusesAnalyses, Durack2014QuantifyingWarming}. Moreover, numerous studies have highlighted the importance of resolving submesoscale dynamics to account for the majority of vertical ocean transport, which is vital for carbon export, fisheries, nutrient availability, and pollution displacement \citep{Pascual2017AAlborEx}. These challenges underscore the need for high-resolution, three-dimensional representations of the ocean state. High-resolution numerical models and data assimilation techniques, which align model outputs with actual observations, are currently the most common solutions \citep{Carrassi2018DataPerspectives,Mourre2004AssimilationBathymetry}.

Operational simulations now assimilate near-real-time observations, including in-situ (ship-based observations, underwater gliders, and floats) and remote sensing data. CIT Satellite observations provide frequent global snapshots of the sea surface, for instance Sea Surface Temperature and Chlorophyll concentration images offer resolutions as fine as 1 km on a daily basis. In contrast, the current capabilities of remote altimeters are limited to a 200 km wavelength for the global ocean at mid-latitudes and about 130 km for the Mediterranean Sea \citep{Ballarotta2019OnMaps}, though significant advancements are upcoming with the Surface Water and Ocean Topography (SWOT) mission successfully launched in December 2022 \citep{Morrow2019GlobalMission}. Notably, Sea Surface Height (SSH) data are unaffected by cloud cover. However, the uncertainties regarding the ocean interior remain significant due to the sparse distribution of in-situ observations in time and space  \citep{Siegelman2019EnhancedFronts}. As a result, while data-assimilating models adhere to physical balances, they still lack accuracy \citep{Arcucci2021DeepAssimilation}. 

The ocean twin strategy proposes data-driven approaches as a complementary method for revealing the ocean state. In previous oceanographic studies, multivariate methods allowed to elaborate three-dimensional hydrographic fields relying on their vast in-situ measurements collected during ocean campaigns \citep{Cutolo2022DiagnosingApproach, Gomis2001DiagnosticData}. However, these methods are not easily scalable to a global observing system due to the sheer number of parameters involved, such as correlation lengths. Machine learning techniques offer a solution to these scalability issues, as the models are directly learned from the data. A key challenge for these techniques is the need for a substantial quantity of realistic training data. General circulation and process study models play a new role here, providing a cost-effective way to generate large datasets that adhere to ocean physics. Even datasets that only approximately match the true ocean state can be valuable, provided they encompass a wide range of scenarios.This last point is especially crucial in preventing the risk of deep networks memorizing the input climatology rather than capturing the actual ocean dynamics. Such a focus ensures that the networks can understand and adapt to scenarios that significantly deviate from the average, rather than being confined to repetitive patterns. To effectively generalize beyond their training data, neural networks require careful design to preserve relevant input features across their layers. In this context, explainable AI aims to advance beyond the black-box applications typical in ocean remote sensing studies, promoting a deeper understanding of the model workings (\citep{Zhu2017DeepResources}.

Despite these difficulties, recent studies have demonstrated the potential of deep-learning methods for various dynamical system tasks. These range from idealized situations \citep{Fablet2021LearningSolvers} to realistic case studies, such as interpolating missing data in satellite-derived observations of sea surface dynamics \citep{Barth2020DINCAEObservations,Manucharyan2021ATurbulence,Fablet2020JointData}. With regard to reconstructing hydrographic profiles from satellite data, there's a spectrum of approaches: from proof-of-concept studies using self-organizing maps (SOMs) and neural networks (\citet{Charantonis2015CompletionSOM,Gueye2014NeuralParameters}) and feed-forward or long short-term memory (LSTM) neural networks \citep{Contractor2021EfficacyTemperature,Sammartino2020AnObservations,Jiang2021OceanNetwork,Fablet2021LearningSolvers} as well as \citep{Pauthenet2022Four-dimensionalNetworks} relying instead on multilayer perceptron. Even considering these past works the interpolation of temperature and salinity profiles given some in-situ and sea surface information is an open challenge.

In this study, we introduce an innovative modular neural network designed to seamlessly integrate remote-sensing images with in-situ observations for a complete 3D reconstruction of the ocean state. This integration is underpinned by the Optimal Interpolation (OI) scheme's mathematical principles \citep{Gandin1966Objective0d}. Unlike traditional applications of OI, which typically use Euclidean distance to estimate the correlation between points, our approach involves computing distances within a specially designed latent space. A specific module within our neural network transform all our input information into this latent space made of 'clusters'. Within these clusters, non-local correlations become more easily identifiable and can be effectively applied to enhance the correlation matrix. Like attention mechanisms in advanced neural models \citep{Vaswani2017AttentionNeed}, which focus on key aspects in large datasets for tasks such as language processing or image recognition, our neural network module similarly identifies crucial correlational patterns through the latent space of clusters.

We privileged a network structure composed of independent nested modules to facilitate the understanding and analysis of its internal information flow from the input data to the covariance structure. To the best of our knowledge, this is the first work in which neural networks achieve the most optimal combination of remote-sensing and in-situ observations without previous knowledge of the study area's climatology. This study is structured as follows: \autoref{sect:data} presents the main synthetic dataset that we used for the training and testing and some real observations for some preliminar use case scenario. All the details regarding the network architecture are in \autoref{sect:methods} while the results are shown in section 4.

\section{Data}
\label{sect:data}
Neural networks need large amounts of data to be trained appropriately. A common choice in oceanography where such a significant quantity of actual observations are unavailable is relying on numerical models. In our case, we chased NATL60, a simulation based on the Nucleus for European Modelling of the Ocean described. We used the fields of this model to simulate both remote-sensing and in-situ observations in a so-called Observing System Simulation Experiment (OSSE). The model output is sampled in these experiments to replicate the different types of partial observations available. The advantage is that we can quickly check the obtained improvements since the model output also provides the ground truth we aim to reconstruct. The danger of what is usually called called "supervised learning" only aiming to minimize the discrepancy with the provided ground truth is that the network weights memorize the "right answers" so in our context the model climatology. We faced this problem, including two self-supervised terms in our loss function as we describe later but also accurately selecting a highly varying training and test dataset as presented here in \autoref{section:synthdata}.

Finally, we proved the generalization capabilities of our network, testing it with actual multi-platform observations. In particular, we used the remote-sensing products of Sea Surface Temperature (SST) and Chlorophyll-a concentration (CHL) from CMEMS, together with temperature observations from gliders, as described in \autoref{sect:realdata}.

\begin{figure}[htbp]
\begin{center}
\includegraphics[width=\textwidth]{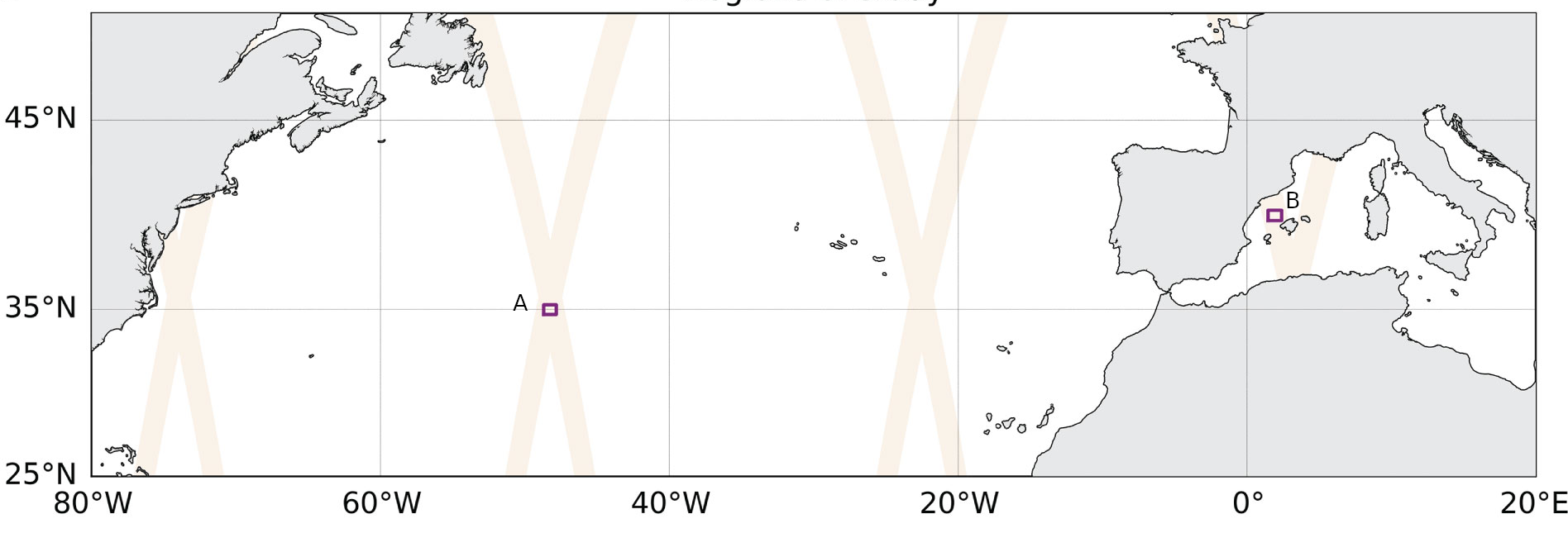}
\end{center}
\caption{Training area (A) and testing area (B) presented with the SWOT passages in the fast-sampling phase.}\label{fig:1}
\end{figure}

\subsection{eNATL60 based OSSE}
\label{section:synthdata}
Our primary experiments utilized the eNATL60 configuration of the Nucleus for the European Modelling of the Ocean (NEMO) model \citep{Gurvan2022NEMOEngine}, featuring a $1/60^{\circ}$ horizontal resolution and 300 vertical levels across the North Atlantic. This high-resolution configuration is essential for understanding ocean dynamics, particularly for surface oceanic motions down to 15 km, which aligns with SWOT observations (\cite{Ajayi2020SpatialModels}). We direct readers to this work for a detailed understanding of NATL60's capabilities. Additionally, numerous studies have employed the non-extended version of NATL60 for resolving fine-scale dynamical processes (\citep{Metref2019ReductionAssimilation,Metref2020Wide-SwathReduction,Fresnay2018ReconstructionAtlantic,Amores2018UpProducts}).

For our training and testing data, we utilized daily averages of Sea Surface Temperature (SST) and Sea Surface Height (SSH), both individually and combined, from the eNATL60 simulation spanning an entire year. Alongside these, we gathered in-situ salinity observations at three specific depths: 5m, 75m, and 150m. Our focus was then to reconstruct the 2D salinity fields at these depths. In particular our analysis predominantly focused on the $5$m and $150$m depths, selected to assess the robustness of our model both within and beyond the mixed-layer depth. To ensure that our network's training and testing in-situ observations mirrored real oceanographic conditions, we adopted two distinct sampling strategies: random and regular. This approach allowed us to evaluate the network's performance in various realistic observational scenarios. The random strategy selects $N$ domain points based on a uniform distribution, while the regular strategy uses a homogeneous grid sampling with a fixed spacing of $\delta x$. By varying $N$ and $\delta x$, we conducted different experiments to observe metric variations.

Our focus was on two marine areas: the subpolar northwest Atlantic for training, and the Western Mediterranean Sea for testing. Both regions are notable for SWOT passages during its rapid-sampling phase (see \autoref{fig:1}). The Mediterranean region, in particular, is known for its dynamic oceanographic characteristics and has been extensively studied through in-situ and remote-sensing methods (
\cite{Ruiz2009MesoscaleData}).
Using different regions for training and testing helps prevent overfitting in the neural network. Overfitting occurs when a model learns the specifics and noise in the training data to an extent that interfere with the model's performance on new data. Since the climatology of the northwest Atlantic differs significantly from that of the Western Mediterranean Sea, we ensure that our network is not just memorizing patterns from the training data but is effectively learning to generalize across different oceanographic contexts. Additionally, we diversified the dataset by sampling the same day with varying $N$ or $\delta x$ values.

For simplicity, our approach assumes a synoptic scenario, where all observations occur simultaneously. Future work will address the non-synoptic nature of actual sampling and explore how the network accommodates this. Furthermore, in this study, we did not incorporate simulated noise or measurement errors into our data, opting to explore these aspects in subsequent research. Despite this, the practical effectiveness of our network is demonstrated through tests using actual observational data, details of which are provided in the following subsection. 

\subsection{Real Observations}
\label{sect:realdata}

\subsubsection{Remote-sensing observations}
In our study, we have used Sea Surface Temperature (SST) and Ocean Color (CHL) imagery from the 18th of February, 2022, distributed by CMEMS. The CHL has $1$ km spatial resolution, and it is a level-3 product obtained by multi-Sensor processing from OceanColor \citep{Volpe2019MediterraneanProcessing}. The SST also has a $1$ km spatial resolution and it is basedd on level-2 product based on multi-channel sea surface temperature (SST) retrievals, which it has generated in real-time from the Infrared Atmospheric Sounding Interferometer (IASI) on the European Meteorological Operational-A (MetOp-A) satellite.

\subsubsection{Glider Observations}
Gliders are autonomous underwater vehicles that allow sustained collection at high spatial resolution ($1$ km) and low costs compared to conventional oceanographic methods. Many studies confirmed the feasibility of using coastal and deep gliders to monitor the spatial and low-frequency variability of the coastal ocean \citep{Zarokanellos2022FrontalObservations,Heslop2012AutonomousSea,Ruiz2019EffectsPhytoplankton,Alvarez2007CombiningOcean}). In this work we used the observations from two gliders in the Balearic Sea as a part of the Calypso 2022 experiment. The two gliders carried out a suite of sensors that measure temperature, conductivity and pressure (CTD), dissolved oxygen (oxygen optode), Chlorophyll fluorescence and Turbidity (FNLTU). The two gliders were programmed to profile from the surface up to 700 m with a vertical speed of 0.18 ± 0.02 m/s and moved horizontally at approximately $20–24$ km per/day. Data were processed following the methodology described in \cite{Troupin2015AManagement}. In this study, we have used the temperature data at 15 m from the 10th of February until the 18th of February.

\section{Methods}
\label{sect:methods}
When sparse observations are available, the most common technique that has been adopted in oceanography and in different fields of science using a gridded product is Optimal Interpolation (OI) \citep{Gandin1966Objective0d}. The technique relies on a solid mathematics basis and has been the state-of-the-art approach for many geophysical products until now. Since the proposed neural approach and specifically our prior builds over the OI framework we reviewed it in \autoref{subsection:OI}. Then, we introduce CLOINet our neural approach and its submodules in \autoref{subsection:CLOINet}. Lastly, we present the metrics we used for bench-marking purposes.

\subsection{Baseline: OINet}
\label{subsection:OI}
A common approach to explain the OI math start considering $\textbf{y}$ as the vector containing all the observations we have of the true state $\textbf{x}$, which is unknown. We can relate them with the following observation model:
\begin{equation}
\textbf{y} = \textbf{H}\textbf{x} + \epsilon
\label{eq:oi_1}
\end{equation}
where $\textbf{H}$ is the observation (or masking) operator, and $\epsilon$ is the observation error. 
Under Gaussian hypotheses for  $\epsilon$ and the prior on $\textbf{x}$, we can obtain the best possible estimation of true state $\textbf{x}_{s}$ given the observations $\textbf{y}$ through a linear operator $\textbf{K}$ (the Kalman gain) (Welch, G., Bishop, G. 1995):
\begin{equation}
\textbf{x}^{s} = \textbf{Ky}
\label{eq:oi_2}
\end{equation}
\begin{equation}
\textbf{K} = \textbf{BH}^{T}(\textbf{HBH}^{T} + \textbf{R})^{-1}
\label{eq:oi_3}
\end{equation}
where $\textbf{R}$ is the observation error covariance matrix and $\textbf{B}$ is the error covariance specific of the analysis. In \autoref{eq:oi_3}, we are assuming an a-priori knowledge of both $\textbf{R}$ and $\textbf{B}$, which could be theoretically obtained by repeating the same experiments many times. Practically, a parameterized covariance matrix is often used to substitute the complete climatology covariances (Wu, 1995; Gaspari  Cohn, 1999). The most common parametrization for this matrix is a Gaussian-shaped correlation, depending only on the points' distances and pre-determined correlation lengths. So for two generic position vectors $r_i$ and $r_j$, we have:
\begin{equation}
\textbf{B}_{i,j} = cov(r_i,r_j) = e^{-\sum_{n=1}^3\frac{(r_{i,n}-r_{j,n})^2}{2c_n^2}}
\label{eq:oi_cov}
\end{equation}
where the sum for dimension $n$ considers the squared difference of the components of the position vectors $r_{i,n}$ and $r_{j,n}$ divided by the squared nth correlation length $c_n$. Regarding the observation error matrix, we asssume from now on that it is diagonal:
\begin{equation}
\textbf{R}_{i,j} = \textbf{I}_{i,j}\epsilon
\label{eq:oi_cov2}
\end{equation}

A different case where the observation errors are correlated is possible. However, $\textbf{R}$ is often assumed diagonal to reduce computational costs \citep{Miyoshi2013AFilter}. Finally, inserting $\textbf{B}$ and $\textbf{R}$ in \autoref{eq:oi_3}
and then in \autoref{eq:oi_2} we can compute our estimated field $\textbf{x}^{s}$.

In our experiments, we established a baseline method with OINet, a simple neural network, that automatically discover the OI correlation lengths among different variables (SST, SSH, and salinity) and dimensions. OINet is then provided with the same input data as CLOINet, including surface fields (SST and/or SSH) and in-situ salinity observations. It operates in a two-step process: the first step involves transforming the multivariate surface fields into a unified field, making it compatible for being used with the salinity observations. The second step is to estimate the three correlation lengths specific to the current set of observations. While the first step involves 2 convolutional layers the second one is a simple feed-forward neural networks able to process a generic number of $O$ observations (see the bottom part of \autoref{fig:scheme}).

Beyond the parameter estimation this module is simply realizing an OI using the formula in $\autoref{eq:oi_cov}$ to calculate the covariances. Notably this approach not only automates the tuning of parameters but also leverages GPU power for more efficient interpolation computations

\subsection{CLuster enhanced Optimal Interpolation Network}
\label{subsection:CLOINet}
Ocean dynamics often display non-local and anisotropic patterns, which traditional Optimal Interpolation (OI) methods struggle to account for effectively. The main challenge with OI lies in its correlation function assumptions, which may not accurately reflect the actual physical conditions of the ocean. For instance, as seen in \autoref{eq:oi_cov} OI typically presumes that points in close proximity are strongly correlated, while distant points are not. However, oceanographic phenomena can exhibit the opposite behavior. For example, ocean fronts, characterized by narrow zones with strong horizontal density gradients, act as boundaries between water masses with distinct physical and bio-optical properties. Conversely, in dynamic ocean features like meanders and eddies, water masses can remain similar over vast distances.

Here, we aim to benefit from the wealth of information from remote sensing regarding the shape of the ocean features, whether they belong to the mesoscale or the submesoscale. The key idea is grouping a set of objects in such a way that each object is more similar to the objects belonging to its same group (called a cluster) than the rest. This procedure in statistics is called clustering. Applying this concept to reconstructing the ocean state, our approach is to reveal non-local correlations by clustering grid points that are part of the same oceanic features. This led us to develop CLOINet (Cluster-enhanced Optimal Interpolation Net), an end-to-end system designed to optimally interpolate sparse in-situ observations using available remote-sensing images. CLOINet is able to process any kind of surface fields (2D images) and in-situ observations (2D masks and observation values). Its main submodule is CLuNet, which transforms 2D fields into fuzzy clusters. While satellite images could directly been passed to this module in-situ observation profiles are initially processed by OINet, which serves as a prior, converting them into images. Finally a further submodule, RefiNet, module merges the fuzzy clusters from both surface fields and observation priors into a final cluster set. Within this latent cluster space an alternative distance could replace the euclidean distance allowing a better estimation of $\textbf{B}$ and consequently obtains the reconstructed field $\textbf{x}^{s}$. 

\begin{figure}[htbp]
\begin{center}
\includegraphics[width=15cm]{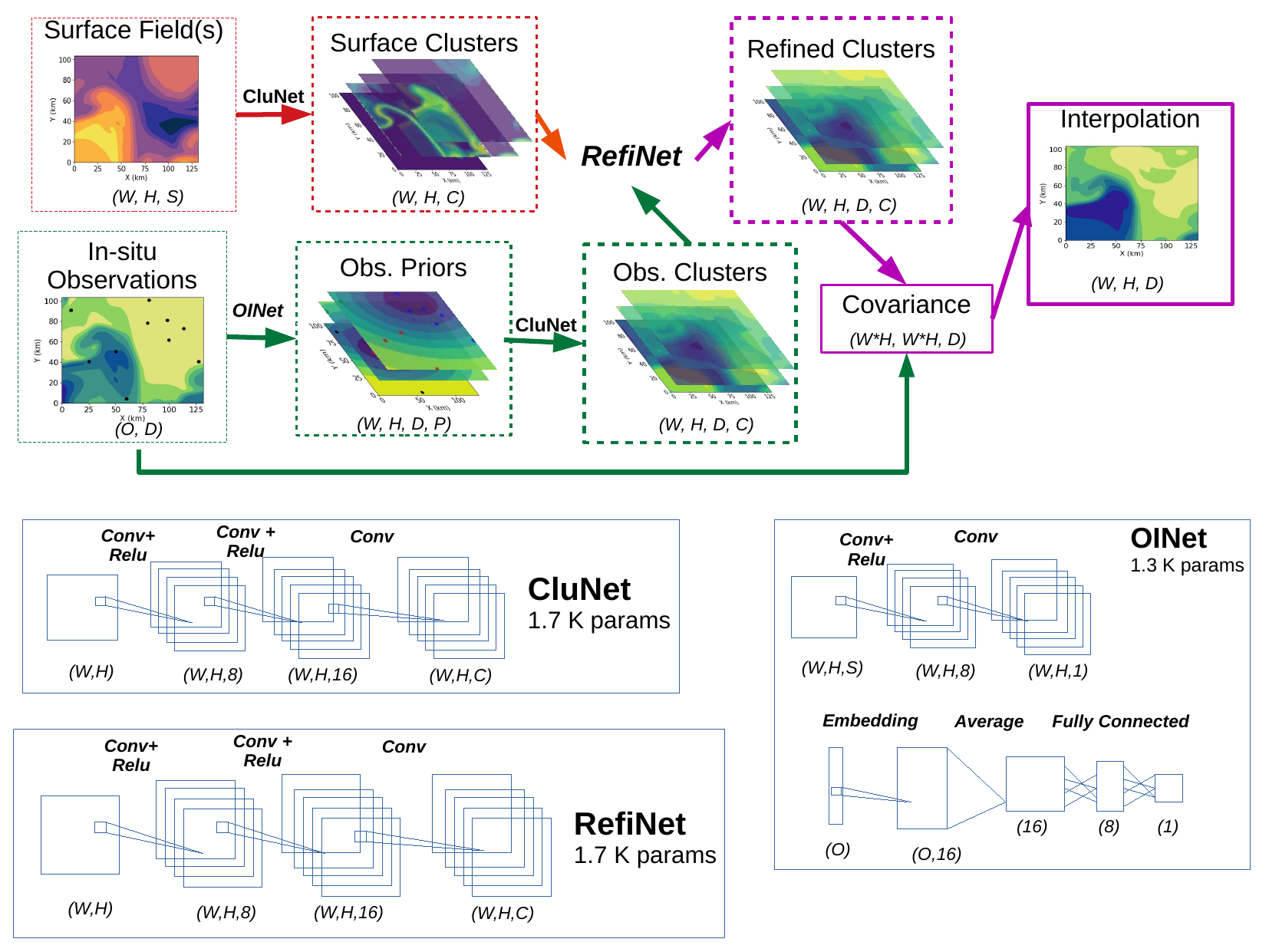}
\end{center}
\caption{Flow chart of CLOINet information processing: Red and green elements (boxes and arrows) represent the processing paths for the SST input surface field and the in-situ salinity observations, respectively. Purple elements indicate the combined use of both inputs. The bold text along the arrows specifies the network module in operation. The line style and width of each box vary to represent different processing stages, ranging from thin and dashed for inputs to solid and thick for outputs. Inside each box, capital letters denote the corresponding tensor dimensions: W for field width, H for field height, S for the number of input surface fields, P for the number of OINet priors, C for the number of clusters (consistent across all modules in our tests), D for the number of depths, and O for the number of in-situ observations. Colorbars are omitted for clarity. The lower part of the image illustrates the CNN architecture of the three modules, along with the number of parameters used.}\label{fig:scheme}
\end{figure}

Our network structure allows a joint training of all modules, minimizing their specific loss function terms summed up in a global loss function.  Convolutional Neural Networks (CNN) layers. Here following, we describe the details of the network submodules and how we obtained the interpolation in an end-to-end scheme also summarized in \autoref{fig:scheme}).

\subsubsection{Clusters space transformation: CluNet}
The first module of our scheme, called CluNet is in charge of transform any images into a set of clusters. Piratically speaking it segments 2D images (like the remote-sensing one or the observations priors) into $C$ clusters of similar points. In this context, we consider two points similar according to their positions (as in \autoref{eq:oi_cov}) but also their values in the input 2D fields. In particular, we worked within the so-called "fuzzy logic", where the membership function $m_{jk}$, which expresses how much the $j$ point belongs to the $k$ cluster, could assume every value between $0$ and $1$. Considering this continuous range means that each grid point could be part of more than one cluster as long as the following normalization holds:
\begin{equation}
\sum_{k=1}^C m_{jk} = 1 \quad \forall j
\label{eq:norm}
\end{equation}
For its non-binary logic, this clustering technique is called "Fuzzy Clustering" or soft k-means. being this last the simpler binary case in which $m_{jk}$ could be just $1$ or $0$. CluNet takes the remote-sensing images as input and gives the tensor composed by all the $m_{jk}$ through various CNN and finally a softmax layer to guarantee \autoref{eq:norm}. 
The associated training loss, referred to as a  Robust fuzzy C-means \citep{Chen2021LearningNetworks} loss, is composed of two terms:
\begin{equation}
\mathcal{L}_{R F C M}(\mathbf{y} ; \theta) =\sum_{j \in \Omega} \sum_{k=1}^C m_{jk}^q(\mathbf{y};\mathbf{\theta})\left\|y_j-v_k\right\|^2 \\
+\beta \sum_{j \in \Omega} \sum_{k=1}^C m_{j k}^q(\mathbf{y} ; \theta) \sum_{l \in N_j m \in M_k} \sum_{l m}m_{lm}^q(\mathbf{y} ; \theta)
\label{eq:rfcm}
\end{equation}
$\mathbf{y}$ is the vector containing the surface field that we want to cluster with $y_j$ its value at point $j$  in our domain $\Omega$. $q$ is a parameter that satisfies $q\geq1$ and controls the amount of fuzzy overlap between clusters. Minimizing the first term achieves that points with high membership function for the $k$ cluster should be similar to its center $v_k$ defined as follows:
\begin{equation}
v_k=\frac{\sum_{j \in \Omega} m_{j k}^q(\mathbf{y} ; \theta) y_j}{\sum_{j \in \Omega} m_{j k}^q(\mathbf{y} ; \theta)}
\end{equation}
The second term  guarantees the membership function's spatial smoothness, forcing the $j$ point to have a similar value to its neighborhood $N_j$. The parameter $\beta$ controls the intensity of this constraint. 

In summary, to obtain the clustering, we minimize \autoref{eq:rfcm} with respect to the parameters of the CNN layers included in CluNet, which stand in the $\mathbf{\theta}$ vector. Since in this loss term, we do not directly provide any ground truth (i.e., the best way of clustering the inputs), this part of the network could be considered self-supervised since it learns indirectly from the rest of the loss term. As it show in \autoref{fig:scheme} we used this module twice, firstly for clustering surface input fields and secondly for clustering the 2D fields coming from the observations priors described hereafter. Consequently in the global loss there are two terms like \autoref{eq:rfcm}.

\subsubsection{Observations Priors}
We have outlined the process by which CluNet segments any set of 2D fields into distinct clusters. To handle in-situ observations, which are essentially vectors of observations at different depths, we utilize OINet to convert them into a series of images that can then be clustered. As previously mentioned, OINet has the capability to autonomously determine the appropriate correlation lengths for a given set of observations and then perform a canonical Optimal Interpolation (OI). In our approach, we generate four different versions of these interpolations, each initiated with correlation lengths that are submultiples of the domain sizes. These parameters, among others, are then fine-tuned during the learning phase. This process results in four fields that CluNet subsequently clusters into areas exhibiting similar values, despite being derived using different correlation lengths. The clusters formed from these observations provide insights into the certainty we have about specific regions and the extent to which a particular depth is influenced by surface conditions. Essentially, this method allows us to address potential anisotropy in the uncertainties without having to rely on fixed length scales. 

\subsubsection{Data fusion in the clusters space: RefiNet}
We now have a set of clusters derived from the surface fields, and an additional set for each depth of the in-situ observations. For each depth, the corresponding sets of surface and observation clusters are processed through RefiNet. The resulting clusters, along with their membership vectors, are used to compute the covariance matrix as follows:
\begin{equation}
\textbf{B}_{i,j} = cov(r_i,r_j) = 1-\sum_{k=1}^C(m'_{ik}-m'_{jk})^2
\label{eq:eoi_cov}
\end{equation}
In this equation, we sum the differences in the membership functions of points ii and jj across all clusters. This process, while bearing similarities to \autoref{eq:oi_cov} deviates by using subtraction instead of an exponential function since $m_{ik}$ and $m_{jk}$ are already bounded within the 0-1 range. This summation represents a non-local distance in the cluster space, replacing the classical Euclidean distance. Consequently, two points within the same cluster (i.e., with similar membership vectors) will be correlated, regardless of their spatial distance.

Using parametrization (\ref{eq:eoi_cov}), we then compute the associated optimal interpolation as
\autoref{eq:oi_3} and then \autoref{eq:oi_2}. This forms an end-to-end architecture that uses remote sensing images and in-situ data to output regularly-gridded vertical profiles (see \autoref{fig:scheme} for the data flow). 

The training loss combines three components: two clustering-based losses \autoref{eq:rfcm} (one for the surface fields and one for the observations priors) and a supervised reconstruction term. So globally we minimize:
\begin{equation}
\mathcal{L} = \alpha\mathcal{L}_{srf_R F C M} + \beta\mathcal{L}_{obs_R F C M} +  \gamma\mathcal{L}_{MSE}
\label{eq:ltot}
\end{equation}
where $\alpha$, $\beta$ and $\gamma$ are the weights of the three loss terms, and $\mathcal{L}_{MSE}$ is given by:
\begin{equation}
\mathcal{L}_{MSE} = (\textbf{x}^{s} - \textbf{x})^2
\label{eq:lmse}
\end{equation}
This last term is just the mean squared error with respect to the ground truth $\textbf{x}$. Within the considered supervised training strategy, the rfcm self-supervised losses \autoref{eq:rfcm} act as regularization terms to improve generalization performance and explainability.
We maintain equal weights of $\alpha$, $\beta$, and $\gamma$ at $1$, as no significant differences were observed with other values. Our network also shows relative insensitivity to other hyperparameters, such as the number of clusters. However, our cross-validation tests indicated that setting this number to $20$ yielded the best results

\subsection{Performance Metrics}
To understand how the clusters sets were changing according to the input data we computed the associated entropy fields. In fact, given that the membership vector is normalized and thus it can be seen as a distribution, its entropy definition is:
\begin{equation}
S_{i}=-\sum_{k=1}^C m'_{ik}\log m'_{ik}
\end{equation}
To assess the performance of the proposed approach, we first define the error between the ground truth and the estimated field value:
\begin{equation}
\textbf{x}_{err}=\textbf{x}-\textbf{x}_{s}
\end{equation}
then we easily obtain our first performance metric: the Root Mean Squared Error (RMSE):
\begin{equation}
RMSE=\sqrt{x_{err}^2}
\end{equation}
We will present this metric in percentage of the standard devation of the ground truth fields. Now considering the standard deviation of the error over the whole $N$ snapshot:
\begin{equation}
\sigma_{err}=\frac{\sum_{t=1}^N\left(\textbf{x}_{err}(t)-\overline{\textbf{x}_{err}(t)}\right)^2}{N}
\end{equation}
we can compute the explained variance score dividing by the standard deviation of the ground truth
\begin{equation}
\sigma_S(x, y)=1-\frac{\sigma_{err}}{\sigma_{true}}
\end{equation}

To highlight the effective resolution of the different reconstruction methods we use the noise-to-signal ratio NSR \citep{Ballarotta2019OnMaps}:
\begin{equation}
NSR(\lambda)=\frac{PSD(\textbf{x}_{err},\lambda)}{PSD(\textbf{x},\lambda)}
\end{equation}
the effective resolution is in fact given by the wavelength $\lambda_s$ where the NSR($\lambda_s$) is $0.5$.

\section{Results and Discussion}
This section first reports numerical experiments using NATL60 OSSEs to evaluate the proposed approach quantitatively. The concluding subsection presents an application to real observations.

\subsection{Clusters Entropy}
\begin{figure}[htbp]
\begin{center}
\includegraphics[width=\textwidth]{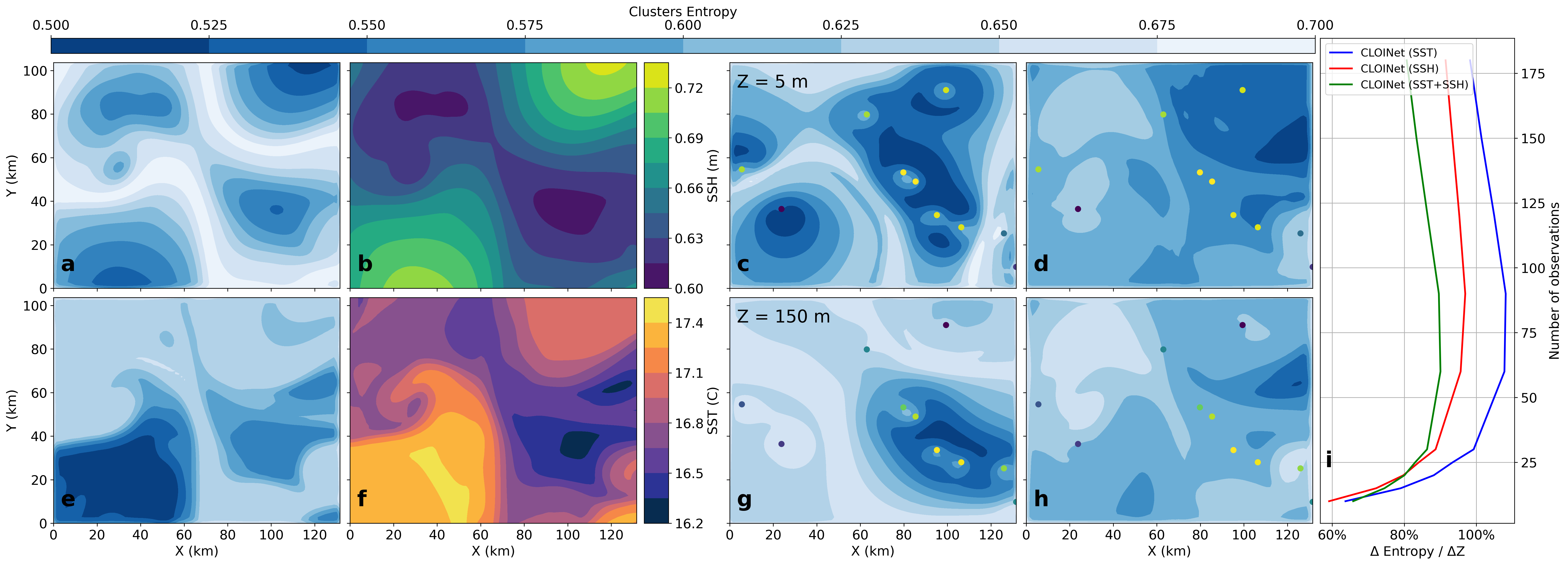}
\end{center}
\caption{The entropy of the cluster sets, resulting from the input SSH and SST, is depicted in panels a and e, respectively, while the corresponding fields themselves are shown in panels b and f. Panels c and d (and g and h) display the entropy of the observation and refined cluster sets at a depth of Z = 5m and Z = 150m, respectively. The dots in these panels represent salinity observations at these depths, with their colors indicating the magnitude of salinity (scale not shown). Panel i illustrates the variation in the entropy of the refined cluster sets along the vertical axis, corresponding to different numbers of observations. The varying colors in this panel represent different networks.}\label{fig:3}
\end{figure}

The initial part of our analysis focuses on understanding how CLOINet, via CluNet and subsequently RefiNet, organizes clusters based on different data inputs: SST, SSH, and various sets of randomly located in-situ salinity observations. To illustrate this, we plotted some example entropy fields in \autoref{fig:3} along with statistics on how entropy changes with an increasing number of observations $N$. In the four panels on the left side of \autoref{fig:3} we display two clusters' entropy fields (panels a and e) and their corresponding input fields for SSH (panel b) and SST (panel f) from a selected snapshot. In the four panels on the right side, we present the entropy associated with the in-situ observations' clusters at two different depths $z=5$ (panel c) and $z=150m$ (panel g) together with the correspondent refined clusters entropy (panel d and h) along with the refined clusters' entropy (panels d and h). 

This particular snapshot was chosen for its submesoscale features. The differences between SSH and SST-based entropy are noticeable; the SSH clusters highlight more prominent features, while SST forms smaller clusters that extend to deeper depths. The correspondence between the surface fields and their cluster entropy is relatively straightforward, but differences in other sets are more subtle. For observation clusters' entropy, we observe lower entropy (blue regions) near points with similar observations. Areas of higher entropy occur between two observation points with differing values. This behavior varies at different depths, explaining the differences between panels c and d. The refined clusters, influenced by both observations and surface fields, exhibit subtler changes, but we can still see an increase in entropy with depth, particularly noticeable in the northeast region of panels d and h.

Beyond this specific snapshot, panel i shows the percentage change in entropy between the two depths, averaged across the entire test dataset as a function of the number of observations. When only SST data is available, the changes in clusters are more pronounced, as SST information is less directly related to the ocean's interior compared to SSH or combined SST and SSH data. As expected, all deltas increase with the number of observations, eventually reaching a saturation point where they decrease. This occurs because the clusters' information becomes less critical, and the field can be reconstructed relying primarily on in-situ observations.

\subsection{RMSE and Correlation}
\begin{figure}[htbp]
\begin{center}
\includegraphics[width=15cm]{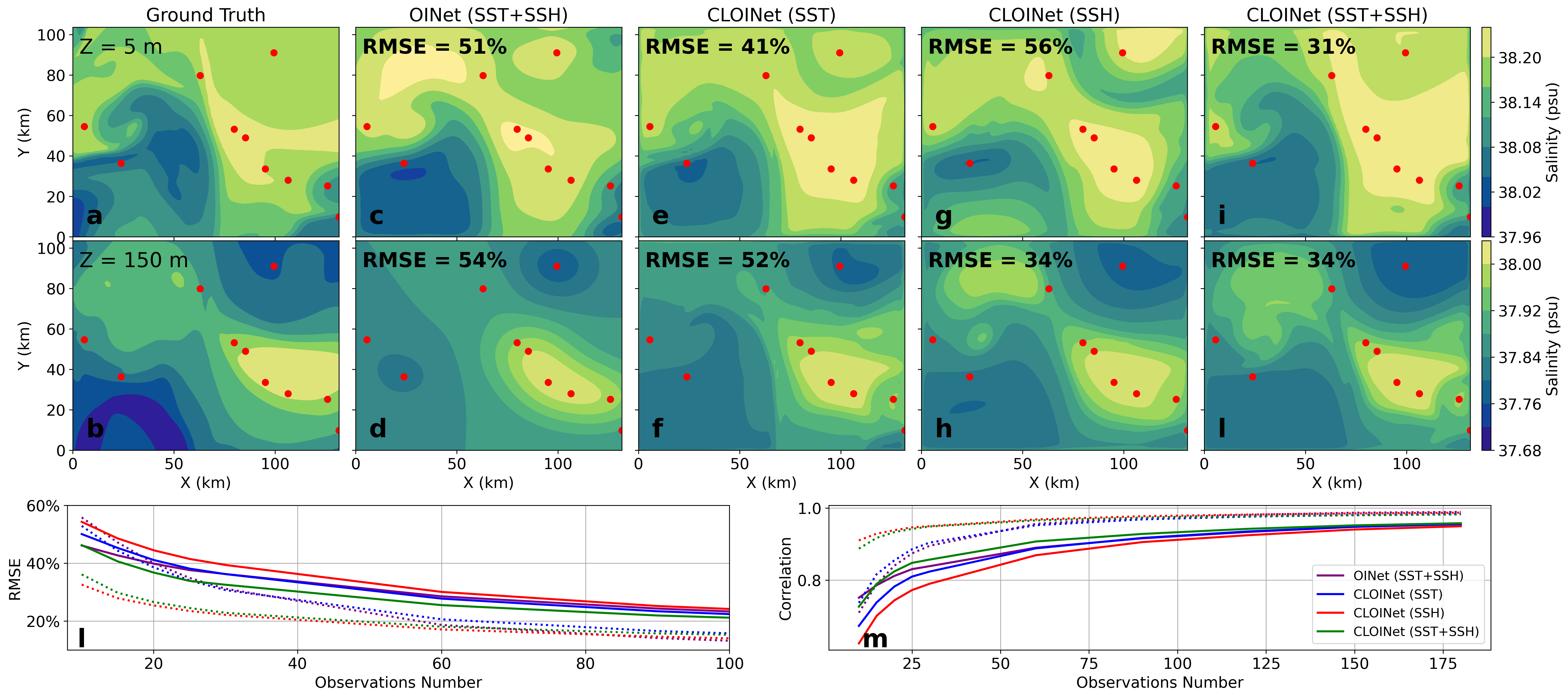}
\end{center}
\caption{Example of salinity field interpolation at Z = 5m (first row) and Z = 150m (second row) with ten random observations. The first column shows the ground truth, and the subsequent columns represent various interpolation methods. The two bottom plots display the RMSE (left) and correlation coefficient (right) as functions of the number of observations in a random sampling scenario, averaged across the entire test dataset. In these plots, the solid line corresponds to Z = 5m, while the dashed line represents Z = 150m.}
\label{fig:4}
\end{figure}

WWe present the outcomes of the random sampling OSSE in \autoref{fig:4}. The first two rows illustrate a ground truth salinity example at two different depths, alongside the reconstructions by the baseline OINet and CLOINet with various surface input fields. Again, we chose the same snapshot from \autoref{fig:3} for its distinct submesoscale features.

OINet can effectively use surface information for reconstructing the surface layer, but it struggles to propagate this information to deeper layers. We also experimented forcinga bigger correlation length in the $z$ axis but we ended up with a reversed scenario: a well-reconstructed bottom layer but a poorly reconstructed  (not shown). This limitation arises because the simple network cannot determine which surface fields to prioritize based on the in-situ observations.

In the case of CLOINet, we observe different results based on the input fields provided. SST leads to better surface interpolations, while SSH is more effective for deeper fields. This outcome aligns with our expectations, as SSH data is depth-integrated and thus more informative than SST for understanding the shape of water masses at depth. Notably, when both SST and SSH are used as inputs, the network effectively leverages their shape information to enhance both surface and interior reconstructions, leading to a reduction in RMSE by about $40\%$ at both depths. 

The results across the entire testing set show similar patterns. On panel l (m) we show how for all methods, the RMSE (correlation) decreases (increases) in proportion to the number of observations. In these plots, solid lines represent surface salinity fields, while dashed lines indicate interior fields at $z=150m$. Interestingly, on average, OINet's performance is comparable to CLOINet's for surface reconstructions but falls short for interior reconstructions. This fact is mostly related with presence or not of submesoscale features as the next subsection analysis will show. The variation in CLOINet's surface inputs shows minimal impact on surface results, with only slight improvements observed in the SST+SSH case. However, the introduction of the SSH field significantly enhances the interior field reconstructions. Once again, this confirms that SSH provides more comprehensive information about the entire water column compared to SST.

\subsection{Resolved scales}
\begin{figure}[htbp]
\begin{center}
\includegraphics[width=15cm]{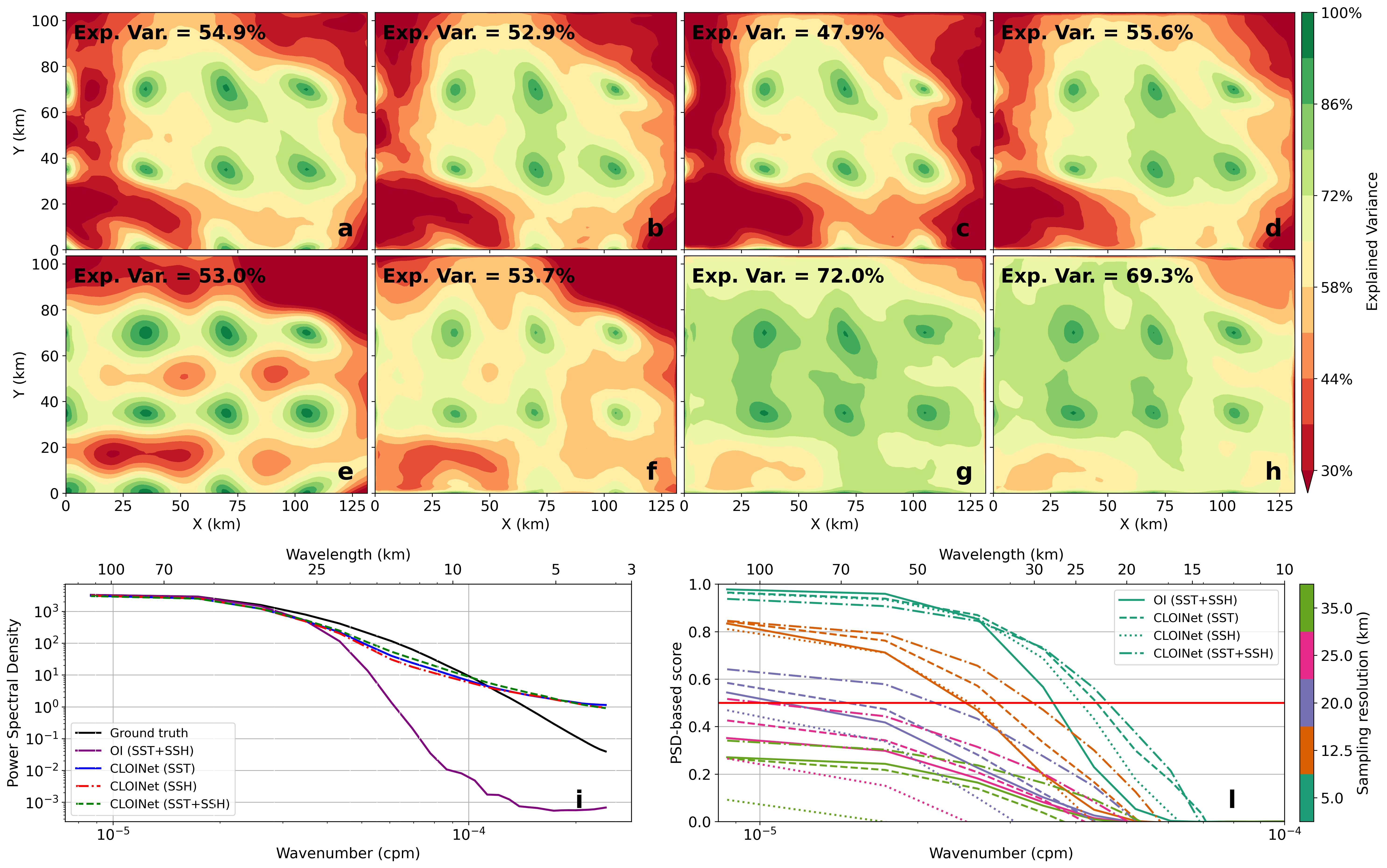}
\end{center}
\caption{This figure shows the surface explained variance for various interpolation methods averaged across the entire test dataset, in a scenario with regular sampling at 45 km intervals. Panels (a), (b), (c), and (d) display the results for OI, CLOINet-SST, CLOINet-SSH, and CLOINet-SST+SSH, respectively. Black dots mark the locations of in-situ observations, and the spatial average value for each method is indicated on the corresponding subplot. Panel (e) compares the Power Spectral Density (PSD) of the different reconstruction methods with the ground truth, with each color representing a different method. Panel (f) shows the corresponding score, where the colors denote different sampling resolutions. In both panels (e) and (f), solid lines represent surface fields, while dashed lines correspond to fields at a depth of z = 150m. The red line across panel (f) marks the 0.5 threshold value, indicating the effective resolution of the interpolation methods.}\label{fig:5}
\end{figure}

In \autoref{fig:5} , we present the results of the OSSE conducted with regular grid sampling, varying the spacing between observations to understand how different methods resolve various spatial scales. Specifically, we examined the impact of sampling resolution on the explained variance and the Power Spectral Density (PSD)-based score. The explained variance for the different reconstruction methods at a sampling resolution of $20$ km is shown in panels a, c, e, and g (and panels b, d, f, and h for the interior field). When provided with the same inputs as OINet (SST and SSH), CLOINet slightly surpasses it on the surface and by about $20\%$ in the interior. Again, we observe superior performance from CLOINet-SSH in the interior, while the inferior performance of the network relying solely on SST suggests that, on average, this field does not significantly account for salinity variability.

The PSD-based score, shown in panel l, indicates the effective resolution of the reconstruction (the point at which the score falls below $0.5$) demostrating how CLOINet generally resolves smaller scales than OINet across various sampling resolutions. For higher resolutions, such as $5$ and $12$ km, CLOINet resolves scales approximately $1.5$ times larger than OINet. The training set's averaged spectra, depicted in panel i, reveal that OINet is typically limited to reconstructing larger scales. Indirectly, this suggests that the variability explained in the test region is predominantly due to larger scales, which even OINet can adequately account for. 

\subsection{Real Ocean Data Preliminary Tests}
\begin{figure}[htbp]
\begin{center}
\includegraphics[width=15cm]{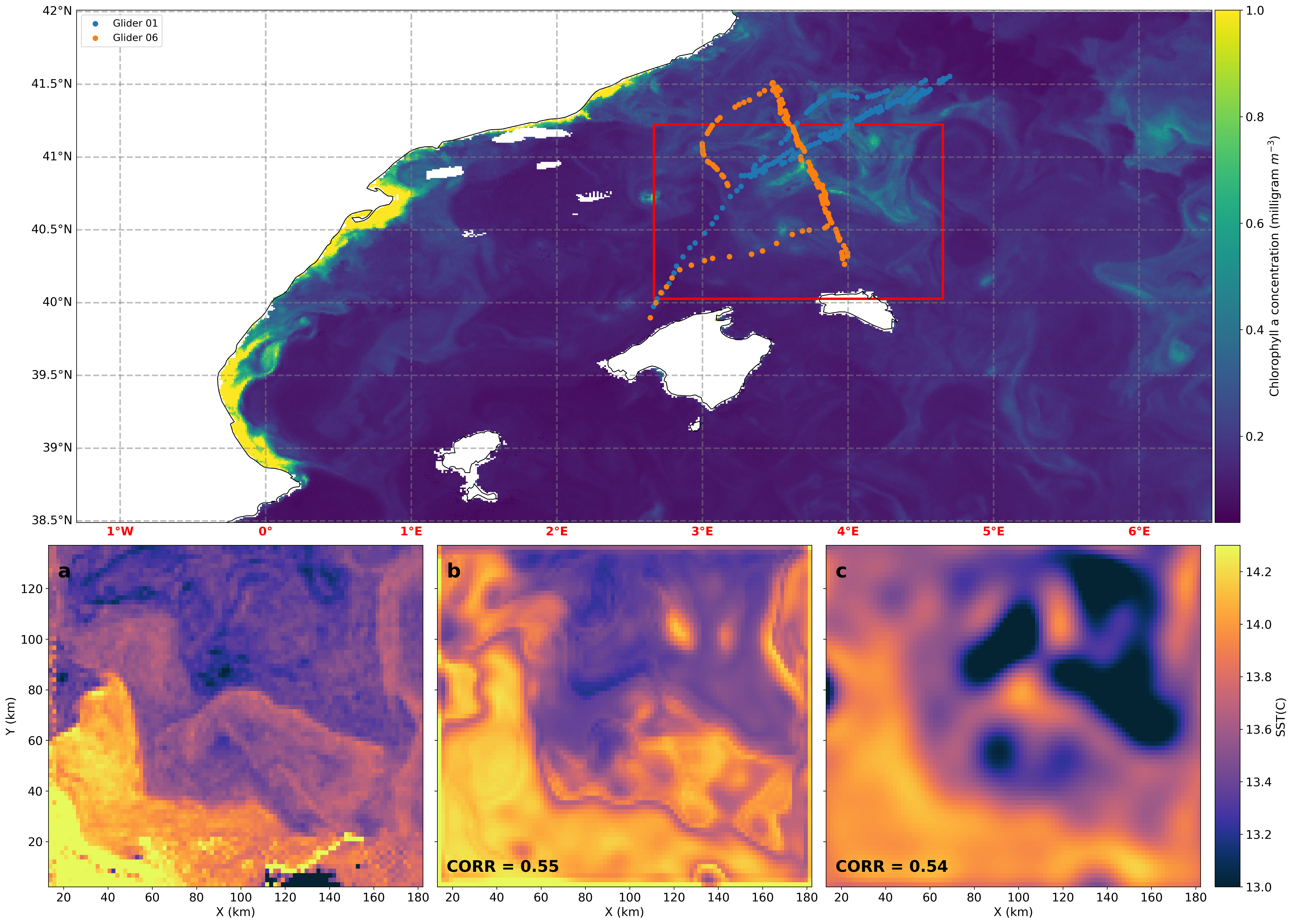}
\end{center}
\caption{The top panel displays the Chlorophyll-a (CHL-a) concentration in the Western Mediterranean Sea on February 18, 2022. The dots represent temperature observations from two gliders over the preceding $48$ hours, while the red box outlines the area where interpolation was performed. The bottom panels, labeled a, b, and c, show the actual Sea Surface Temperature (SST) this same day and the reconstructed SST using CLOINet and OINet, respectively. The correlation values for each reconstruction method are indicated in the corresponding plots.}\label{fig:6}
\end{figure}

In line with many deep learning studies, our research focuses on applying neural networks, initially trained on synthetic data, to real-world observations. We evaluated CLOINet's effectiveness in improving Sea Surface Temperature (SST) estimates using glider surface temperature observations, enhanced with shape information from a Chlorophyll (CHL) snapshot (refer to \autoref{fig:6}). Both OINet and CLOINet were able to reconstruct the general SST pattern observed in reality. CLOINet demonstrated a slightly superior performance, as evidenced by higher correlation values. This improvement aligns with our qualitative observations, suggesting that CLOINet more accurately preserves submesoscale features. Notably, this achievement was realized without the networks being specifically trained on CHL data. In these preliminary tests, the CHL data was provided as if it were the SST and SSH fields, demonstrating the networks' versatility in utilizing shape information from various types of variables. Achieving similar levels of accuracy with traditional Optimal Interpolation (OI) methods would be more complex, likely necessitating intricate, predefined multi-variate correlation functions and extensive parameter tuning.

\section{Conclusion}
In this, we presented CLOINet, a comprehensive end-to-end neural network designed to interpolate sparse in-situ observations into a full 3D field leveraging shape information from various ocean remote sensing imagery types. We conducted end-to-end training of CLOINet within a supervised framework, using Observing System Simulation Experiments (OSSEs) based on the NEMO-derived NATL60 simulation. Our study focuses on comparing the reconstruction capabilities of CLOINet with those of a data-driven version of classical Optimal Interpolation, which we have named OINet. This comparison also extends to applications involving real observational data.

Our research covered various scenarios, including both randomly and regularly spaced in-situ salinity observations, paired with different remote sensing inputs such as Sea Surface Temperature (SST), Sea Surface Height (SSH), or a combination of both. Upon creating a 3D salinity field, we thoroughly analyzed how our performance metrics responded to variations in the number and density of in-situ observations.

In dense regular sampling we showed how CLOINet was able to resolve scales $1.5$ smaller scales compared to OINet while in random sampling contexts, CLOINet showed enhanced performance in terms of both RMSE and correlation, especially notable when limited observations were available. This improvement was significant in scenarios involving in-depth fields and areas rich in submesoscale features, where RMSE improvements reached as high as $40\%$.

Despite not incorporating simulated errors to mimic actual sampling instruments, the promising results with real data highlight the potential of our approach in operational contexts. Both OINet and CLOINet adeptly handled noisy CHL fields and successfully reconstructed the general pattern of an unseen SST field, without specific training for this task.

Our training approach, which combined two self-supervised losses with a supervised reconstruction loss, enabled the network to generalize effectively. This was evident as it performed accurately in the Western Mediterranean test area, distinctly different from the North Atlantic training region. This suggests that our method is not limited by specific regional climatology and could potentially be scaled for global application.

Overall, the modular design of CLOINet not only enhances our understanding of its internal processes but also positions it for future enhancements. One promising direction for subsequent research is extending the model to incorporate space-time dynamics. Another intriguing possibility is employing this neural network approach for guiding an adaptive sampling multi-platform ocean campaign. Given the significant role of SSH data in assessing the reconstruction of the deeper water layers, the upcoming high-resolution SSH observations from SWOT present an exciting opportunity for further refining and applying CLOINet.

\bibliographystyle{unsrtnat}
\bibliography{new_ref}  

\begin{thebibliography}{40}
\providecommand{\natexlab}[1]{#1}
\providecommand{\url}[1]{\texttt{#1}}
\expandafter\ifx\csname urlstyle\endcsname\relax
  \providecommand{\doi}[1]{doi: #1}\else
  \providecommand{\doi}{doi: \begingroup \urlstyle{rm}\Url}\fi

\bibitem[Pascual et~al.(2021)Pascual, Mac{\'{i}}as, LastNameLastNameLastNameLastNameLastNameLastNameLastNameTintor{\'{e}}, Turiel, Ballabrera-Poy, Castro, LastNameLastNameLastNameLastNameLastNameLastNameLastNameMarb{\`{a}}, Coll, Dachs, Huertas, Sallar{\`{e}}s, Gonz{\'{a}}lez, Tovar-S{\'{a}}nchez, LastNameLastNameLastNameLastNameLastNameLastNameLastNameGabarr{\'{o}}, Ruiz~Segura, Orfila, Logares, Al{\'{o}}s, Pintado, and Crespo~Solana]{Pascual2021White2030}
Ananda Pascual, Diego Mac{\'{i}}as, Joaquín LastNameLastNameLastNameLastNameLastNameLastNameLastNameTintor{\'{e}}, Antonio Turiel, Joaquim Ballabrera-Poy, Carmen~G. Castro, Núria LastNameLastNameLastNameLastNameLastNameLastNameLastNameMarb{\`{a}}, Marta Coll, Jordi Dachs, I.~Emma Huertas, Valentí Sallar{\`{e}}s, Ángel~F. Gonz{\'{a}}lez, Antonio Tovar-S{\'{a}}nchez, Carolina LastNameLastNameLastNameLastNameLastNameLastNameLastNameGabarr{\'{o}}, Javier Ruiz~Segura, Alejandro Orfila, Ramiro Logares, Josep Al{\'{o}}s, José Pintado, and Ana Crespo~Solana.
\newblock \emph{{White Paper 13: Ocean science challenges for 2030}}.
\newblock Consejo Superior de Investigaciones Cient{\'{i}}ficas (Espa{\~{n}}a), 2021.
\newblock ISBN 978-84-00-10762-8.
\newblock URL \url{https://digital.csic.es/handle/10261/272217}.

\bibitem[Wang et~al.(2018)Wang, Cheng, Abraham, and Li]{Wang2018ConsensusesAnalyses}
Gongjie Wang, Lijing Cheng, John Abraham, and Chongyin Li.
\newblock {Consensuses and discrepancies of basin-scale ocean heat content changes in different ocean analyses}.
\newblock \emph{Climate Dynamics}, 50\penalty0 (7-8):\penalty0 2471--2487, 4 2018.
\newblock ISSN 14320894.
\newblock \doi{10.1007/S00382-017-3751-5/FIGURES/13}.
\newblock URL \url{https://link.springer.com/article/10.1007/s00382-017-3751-5}.

\bibitem[Durack et~al.(2014)Durack, Gleckler, Landerer, and Taylor]{Durack2014QuantifyingWarming}
Paul~J. Durack, Peter~J. Gleckler, Felix~W. Landerer, and Karl~E. Taylor.
\newblock {Quantifying underestimates of long-term upper-ocean warming}.
\newblock \emph{Nature Climate Change}, 4\penalty0 (11):\penalty0 999--1005, 11 2014.
\newblock ISSN 17586798.
\newblock \doi{10.1038/NCLIMATE2389}.

\bibitem[Pascual et~al.(2017)Pascual, Ruiz, Olita, Troupin, Claret, Casas, Mourre, Poulain, Tovar-Sanchez, Capet, Mason, Allen, Mahadevan, and Tintor{\'{e}}]{Pascual2017AAlborEx}
Ananda Pascual, Simon Ruiz, Antonio Olita, Charles Troupin, Mariona Claret, Benjamin Casas, Baptiste Mourre, Pierre~Marie Poulain, Antonio Tovar-Sanchez, Arthur Capet, Evan Mason, John~T. Allen, Amala Mahadevan, and Joaquín Tintor{\'{e}}.
\newblock {A multiplatform experiment to unravel meso- and submesoscale processes in an intense front (AlborEx)}.
\newblock \emph{Frontiers in Marine Science}, 4:\penalty0 39, 2 2017.
\newblock ISSN 22967745.
\newblock \doi{10.3389/FMARS.2017.00039/BIBTEX}.

\bibitem[Carrassi et~al.(2018)Carrassi, Bocquet, Bertino, and Evensen]{Carrassi2018DataPerspectives}
Alberto Carrassi, Marc Bocquet, Laurent Bertino, and Geir Evensen.
\newblock {Data assimilation in the geosciences: An overview of methods, issues, and perspectives}.
\newblock \emph{Wiley Interdisciplinary Reviews: Climate Change}, 9\penalty0 (5):\penalty0 e535, 9 2018.
\newblock ISSN 1757-7799.
\newblock \doi{10.1002/WCC.535}.
\newblock URL \url{https://onlinelibrary.wiley.com/doi/full/10.1002/wcc.535 https://onlinelibrary.wiley.com/doi/abs/10.1002/wcc.535 https://wires.onlinelibrary.wiley.com/doi/10.1002/wcc.535}.

\bibitem[Mourre et~al.(2004)Mourre, De~Mey, Lyard, and Le~Provost]{Mourre2004AssimilationBathymetry}
Baptiste Mourre, Pierre De~Mey, Florent Lyard, and Christian Le~Provost.
\newblock {Assimilation of sea level data over continental shelves: an ensemble method for the exploration of model errors due to uncertainties in bathymetry}.
\newblock \emph{Dynamics of Atmospheres and Oceans}, 38\penalty0 (2):\penalty0 93--121, 11 2004.
\newblock ISSN 0377-0265.
\newblock \doi{10.1016/J.DYNATMOCE.2004.09.001}.

\bibitem[Ballarotta et~al.(2019)Ballarotta, Ubelmann, Pujol, Taburet, Fournier, Legeais, Faug{\`{e}}re, Delepoulle, Chelton, Dibarboure, and Picot]{Ballarotta2019OnMaps}
Maxime Ballarotta, Clément Ubelmann, Marie~Isabelle Pujol, Guillaume Taburet, Florent Fournier, Jean~François Legeais, Yannice Faug{\`{e}}re, Antoine Delepoulle, Dudley Chelton, Gérald Dibarboure, and Nicolas Picot.
\newblock {On the resolutions of ocean altimetry maps}.
\newblock \emph{Ocean Science}, 15\penalty0 (4):\penalty0 1091--1109, 8 2019.
\newblock ISSN 18120792.
\newblock \doi{10.5194/OS-15-1091-2019}.

\bibitem[Morrow et~al.(2019)Morrow, Fu, Ardhuin, Benkiran, Chapron, Cosme, D'Ovidio, Farrar, Gille, Lapeyre, Le~Traon, Pascual, Ponte, Qiu, Rascle, Ubelmann, Wang, and Zaron]{Morrow2019GlobalMission}
Rosemary Morrow, Lee~Lueng Fu, Fabrice Ardhuin, Mounir Benkiran, Bertrand Chapron, Emmanuel Cosme, Francesco D'Ovidio, J.~T. Farrar, Sarah~T. Gille, Guillaume Lapeyre, Pierre~Yves Le~Traon, Ananda Pascual, Aurelien Ponte, Bo~Qiu, Nicolas Rascle, Clement Ubelmann, Jinbo Wang, and Edward Zaron.
\newblock {Global observations of fine-scale ocean surface topography with the Surface Water and Ocean Topography (SWOT) Mission}.
\newblock \emph{Frontiers in Marine Science}, 6\penalty0 (APR):\penalty0 232, 2019.
\newblock ISSN 22967745.
\newblock \doi{10.3389/FMARS.2019.00232/BIBTEX}.

\bibitem[Siegelman et~al.(2019)Siegelman, Klein, Rivi{\`{e}}re, Thompson, Torres, Flexas, and Menemenlis]{Siegelman2019EnhancedFronts}
Lia Siegelman, Patrice Klein, Pascal Rivi{\`{e}}re, Andrew~F. Thompson, Hector~S. Torres, Mar Flexas, and Dimitris Menemenlis.
\newblock {Enhanced upward heat transport at deep submesoscale ocean fronts}.
\newblock \emph{Nature Geoscience 2019 13:1}, 13\penalty0 (1):\penalty0 50--55, 12 2019.
\newblock ISSN 1752-0908.
\newblock \doi{10.1038/s41561-019-0489-1}.
\newblock URL \url{https://www.nature.com/articles/s41561-019-0489-1}.

\bibitem[Arcucci et~al.(2021)Arcucci, Zhu, Hu, and Guo]{Arcucci2021DeepAssimilation}
Rossella Arcucci, Jiangcheng Zhu, Shuang Hu, and Yi~Ke Guo.
\newblock {Deep Data Assimilation: Integrating Deep Learning with Data Assimilation}.
\newblock \emph{Applied Sciences 2021, Vol. 11, Page 1114}, 11\penalty0 (3):\penalty0 1114, 1 2021.
\newblock ISSN 2076-3417.
\newblock \doi{10.3390/APP11031114}.
\newblock URL \url{https://www.mdpi.com/2076-3417/11/3/1114/htm https://www.mdpi.com/2076-3417/11/3/1114}.

\bibitem[Cutolo et~al.(2022)Cutolo, Pascual, Ruiz, Johnston, Freilich, Mahadevan, Shcherbina, Poulain, Ozgokmen, Centurioni, Rudnick, and D’Asaro]{Cutolo2022DiagnosingApproach}
Eugenio Cutolo, Ananda Pascual, Simón Ruiz, T.M.~Shaun Johnston, Mara Freilich, Amala Mahadevan, Andrey Shcherbina, Pierre-Marie Poulain, Tamay Ozgokmen, Luca~R. Centurioni, Daniel~L. Rudnick, and Eric D’Asaro.
\newblock {Diagnosing Frontal Dynamics from Observations using a Variational Approach}.
\newblock \emph{Journal of Geophysical Research: Oceans}, page e2021JC018336, 9 2022.
\newblock ISSN 2169-9291.
\newblock \doi{10.1029/2021JC018336}.
\newblock URL \url{https://onlinelibrary.wiley.com/doi/full/10.1029/2021JC018336 https://onlinelibrary.wiley.com/doi/abs/10.1029/2021JC018336 https://agupubs.onlinelibrary.wiley.com/doi/10.1029/2021JC018336}.

\bibitem[Gomis et~al.(2001)Gomis, Ruiz, and Pedder]{Gomis2001DiagnosticData}
D.~Gomis, S.~Ruiz, and M.~A. Pedder.
\newblock {Diagnostic analysis of the 3D ageostrophic circulation from a multivariate spatial interpolation of CTD and ADCP data}.
\newblock \emph{Deep Sea Research Part I: Oceanographic Research Papers}, 48\penalty0 (1):\penalty0 269--295, 1 2001.
\newblock ISSN 0967-0637.
\newblock \doi{10.1016/S0967-0637(00)00060-1}.

\bibitem[Zhu et~al.(2017)Zhu, Tuia, Mou, Xia, Zhang, Xu, and Fraundorfer]{Zhu2017DeepResources}
Xiao~Xiang Zhu, Devis Tuia, Lichao Mou, Gui~Song Xia, Liangpei Zhang, Feng Xu, and Friedrich Fraundorfer.
\newblock {Deep Learning in Remote Sensing: A Comprehensive Review and List of Resources}.
\newblock \emph{IEEE Geoscience and Remote Sensing Magazine}, 5\penalty0 (4):\penalty0 8--36, 12 2017.
\newblock ISSN 21686831.
\newblock \doi{10.1109/MGRS.2017.2762307}.

\bibitem[Fablet et~al.(2021)Fablet, Chapron, Drumetz, M{\'{e}}min, Pannekoucke, and Rousseau]{Fablet2021LearningSolvers}
R.~Fablet, B.~Chapron, L.~Drumetz, E.~M{\'{e}}min, O.~Pannekoucke, and F.~Rousseau.
\newblock {Learning Variational Data Assimilation Models and Solvers}.
\newblock \emph{Journal of Advances in Modeling Earth Systems}, 13\penalty0 (10):\penalty0 e2021MS002572, 10 2021.
\newblock ISSN 1942-2466.
\newblock \doi{10.1029/2021MS002572}.
\newblock URL \url{https://onlinelibrary.wiley.com/doi/full/10.1029/2021MS002572 https://onlinelibrary.wiley.com/doi/abs/10.1029/2021MS002572 https://agupubs.onlinelibrary.wiley.com/doi/10.1029/2021MS002572}.

\bibitem[Barth et~al.(2020)Barth, Alvera-Azc{\'{a}}rate, Licer, and Beckers]{Barth2020DINCAEObservations}
Alexander Barth, Aida Alvera-Azc{\'{a}}rate, Matjaz Licer, and Jean~Marie Beckers.
\newblock {DINCAE 1.0: A convolutional neural network with error estimates to reconstruct sea surface temperature satellite observations}.
\newblock \emph{Geoscientific Model Development}, 13\penalty0 (3):\penalty0 1609--1622, 3 2020.
\newblock ISSN 19919603.
\newblock \doi{10.5194/GMD-13-1609-2020}.

\bibitem[Manucharyan et~al.(2021)Manucharyan, Siegelman, and Klein]{Manucharyan2021ATurbulence}
Georgy~E. Manucharyan, Lia Siegelman, and Patrice Klein.
\newblock {A Deep Learning Approach to Spatiotemporal Sea Surface Height Interpolation and Estimation of Deep Currents in Geostrophic Ocean Turbulence}.
\newblock \emph{Journal of Advances in Modeling Earth Systems}, 13\penalty0 (1):\penalty0 e2019MS001965, 1 2021.
\newblock ISSN 1942-2466.
\newblock \doi{10.1029/2019MS001965}.
\newblock URL \url{https://onlinelibrary.wiley.com/doi/full/10.1029/2019MS001965 https://onlinelibrary.wiley.com/doi/abs/10.1029/2019MS001965 https://agupubs.onlinelibrary.wiley.com/doi/10.1029/2019MS001965}.

\bibitem[Fablet et~al.(2020)Fablet, Drumetz, and Rousseau]{Fablet2020JointData}
Ronan Fablet, Lucas Drumetz, and Francois Rousseau.
\newblock {Joint learning of variational representations and solvers for inverse problems with partially-observed data}.
\newblock 6 2020.
\newblock \doi{10.48550/arxiv.2006.03653}.
\newblock URL \url{https://arxiv.org/abs/2006.03653v1}.

\bibitem[Charantonis et~al.(2015)Charantonis, Testor, Mortier, D'Ortenzio, and Thiria]{Charantonis2015CompletionSOM}
Anastase~Alexandre Charantonis, Pierre Testor, Laurent Mortier, Fabrizio D'Ortenzio, and Sylvie Thiria.
\newblock {Completion of a Sparse GLIDER Database Using Multi-iterative Self-Organizing Maps (ITCOMP SOM)}.
\newblock \emph{Procedia Computer Science}, 51\penalty0 (1):\penalty0 2198--2206, 1 2015.
\newblock ISSN 1877-0509.
\newblock \doi{10.1016/J.PROCS.2015.05.496}.

\bibitem[Gueye et~al.(2014)Gueye, Niang, Arnault, Thiria, and Cr{\'{e}}pon]{Gueye2014NeuralParameters}
Mbaye~Babacar Gueye, Awa Niang, Sabine Arnault, Sylvie Thiria, and Michel Cr{\'{e}}pon.
\newblock {Neural approach to inverting complex system: Application to ocean salinity profile estimation from surface parameters}.
\newblock \emph{Computers {\&} Geosciences}, 72:\penalty0 201--209, 11 2014.
\newblock ISSN 0098-3004.
\newblock \doi{10.1016/J.CAGEO.2014.07.012}.

\bibitem[Contractor and Roughan(2021)]{Contractor2021EfficacyTemperature}
Steefan Contractor and Moninya Roughan.
\newblock {Efficacy of Feedforward and LSTM Neural Networks at Predicting and Gap Filling Coastal Ocean Timeseries: Oxygen, Nutrients, and Temperature}.
\newblock \emph{Frontiers in Marine Science}, 8:\penalty0 368, 5 2021.
\newblock ISSN 22967745.
\newblock \doi{10.3389/FMARS.2021.637759/BIBTEX}.

\bibitem[Sammartino et~al.(2020)Sammartino, Nardelli, Marullo, and Santoleri]{Sammartino2020AnObservations}
Michela Sammartino, Bruno~Buongiorno Nardelli, Salvatore Marullo, and Rosalia Santoleri.
\newblock {An Artificial Neural Network to Infer the Mediterranean 3D Chlorophyll-a and Temperature Fields from Remote Sensing Observations}.
\newblock \emph{Remote Sensing 2020, Vol. 12, Page 4123}, 12\penalty0 (24):\penalty0 4123, 12 2020.
\newblock ISSN 2072-4292.
\newblock \doi{10.3390/RS12244123}.
\newblock URL \url{https://www.mdpi.com/2072-4292/12/24/4123/htm https://www.mdpi.com/2072-4292/12/24/4123}.

\bibitem[Jiang et~al.(2021)Jiang, Ma, Wang, Shen, and Yuan]{Jiang2021OceanNetwork}
Fan Jiang, Jitong Ma, Baosen Wang, Feifei Shen, and Lingling Yuan.
\newblock {Ocean Observation Data Prediction for Argo Data Quality Control Using Deep Bidirectional LSTM Network}.
\newblock \emph{Security and Communication Networks}, 2021, 2021.
\newblock ISSN 19390122.
\newblock \doi{10.1155/2021/5665386}.

\bibitem[Pauthenet et~al.(2022)Pauthenet, Bachelot, Balem, Maze, Tr{\'{e}}guier, Roquet, Fablet, and Tandeo]{Pauthenet2022Four-dimensionalNetworks}
Etienne Pauthenet, Loïc Bachelot, Kevin Balem, Guillaume Maze, Anne-Marie Tr{\'{e}}guier, Fabien Roquet, Ronan Fablet, and Pierre Tandeo.
\newblock {Four-dimensional temperature, salinity and mixed-layer depth in the Gulf Stream, reconstructed from remote-sensing and in situ observations with neural networks}.
\newblock \emph{Ocean Science}, 18\penalty0 (4):\penalty0 1221--1244, 8 2022.
\newblock ISSN 18120792.
\newblock \doi{10.5194/OS-18-1221-2022}.

\bibitem[Gandin(1966)]{Gandin1966Objective0d}
L.~S. Gandin.
\newblock {Objective analysis of meteorological fields. Translated from the Russian. Jerusalem (Israel Program for Scientific Translations), 1965. Pp. vi, 242: 53 Figures; 28 Tables. {\pounds}4 1s. 0d}.
\newblock \emph{Quarterly Journal of the Royal Meteorological Society}, 92\penalty0 (393):\penalty0 447--447, 7 1966.
\newblock ISSN 1477-870X.
\newblock \doi{10.1002/QJ.49709239320}.
\newblock URL \url{https://onlinelibrary.wiley.com/doi/full/10.1002/qj.49709239320 https://onlinelibrary.wiley.com/doi/abs/10.1002/qj.49709239320 https://rmets.onlinelibrary.wiley.com/doi/10.1002/qj.49709239320}.

\bibitem[Vaswani et~al.(2017)Vaswani, Shazeer, Parmar, Uszkoreit, Jones, Gomez, Kaiser, and Polosukhin]{Vaswani2017AttentionNeed}
Ashish Vaswani, Noam Shazeer, Niki Parmar, Jakob Uszkoreit, Llion Jones, Aidan~N. Gomez, Łukasz Kaiser, and Illia Polosukhin.
\newblock {Attention Is All You Need}.
\newblock \emph{Advances in Neural Information Processing Systems}, 2017-December:\penalty0 5999--6009, 6 2017.
\newblock ISSN 10495258.
\newblock \doi{10.48550/arxiv.1706.03762}.
\newblock URL \url{https://arxiv.org/abs/1706.03762v5}.

\bibitem[Gurvan et~al.(2022)Gurvan, Bourdall{\'{e}}-Badie, Chanut, Clementi, Coward, Eth{\'{e}}, Iovino, Lea, L{\'{e}}vy, Lovato, Martin, Masson, Mocavero, Rousset, Storkey, M{\"{u}}eller, Nurser, Bell, Samson, Mathiot, Mele, and Moulin]{Gurvan2022NEMOEngine}
Madec Gurvan, Romain Bourdall{\'{e}}-Badie, Jérôme Chanut, Emanuela Clementi, Andrew Coward, Christian Eth{\'{e}}, Doroteaciro Iovino, Dan Lea, Claire L{\'{e}}vy, Tomas Lovato, Nicolas Martin, Sébastien Masson, Silvia Mocavero, Clément Rousset, Dave Storkey, Simon M{\"{u}}eller, George Nurser, Mike Bell, Guillaume Samson, Pierre Mathiot, Francesca Mele, and Aimie Moulin.
\newblock {NEMO ocean engine}.
\newblock Technical report, 3 2022.
\newblock URL \url{https://zenodo.org/record/6334656}.

\bibitem[Ajayi et~al.(2020)Ajayi, Le~Sommer, Chassignet, Molines, Xu, Albert, and Cosme]{Ajayi2020SpatialModels}
Adekunle Ajayi, Julien Le~Sommer, Eric Chassignet, Jean~Marc Molines, Xiaobiao Xu, Aurelie Albert, and Emmanuel Cosme.
\newblock {Spatial and Temporal Variability of the North Atlantic Eddy Field From Two Kilometric-Resolution Ocean Models}.
\newblock \emph{Journal of Geophysical Research: Oceans}, 125\penalty0 (5):\penalty0 e2019JC015827, 5 2020.
\newblock ISSN 2169-9291.
\newblock \doi{10.1029/2019JC015827}.
\newblock URL \url{https://onlinelibrary.wiley.com/doi/full/10.1029/2019JC015827 https://onlinelibrary.wiley.com/doi/abs/10.1029/2019JC015827 https://agupubs.onlinelibrary.wiley.com/doi/10.1029/2019JC015827}.

\bibitem[Metref et~al.(2019)Metref, Cosme, Le~Sommer, Poel, Brankart, Verron, and Navarro]{Metref2019ReductionAssimilation}
Sammy Metref, Emmanuel Cosme, Julien Le~Sommer, Nora Poel, Jean~Michel Brankart, Jacques Verron, and Laura~Gómez Navarro.
\newblock {Reduction of Spatially Structured Errors in Wide-Swath Altimetric Satellite Data Using Data Assimilation}.
\newblock \emph{Remote Sensing 2019, Vol. 11, Page 1336}, 11\penalty0 (11):\penalty0 1336, 6 2019.
\newblock ISSN 2072-4292.
\newblock \doi{10.3390/RS11111336}.
\newblock URL \url{https://www.mdpi.com/2072-4292/11/11/1336/htm https://www.mdpi.com/2072-4292/11/11/1336}.

\bibitem[Metref et~al.(2020)Metref, Cosme, Le~Guillou, Le~Sommer, Brankart, and Verron]{Metref2020Wide-SwathReduction}
Sammy Metref, Emmanuel Cosme, Florian Le~Guillou, Julien Le~Sommer, Jean~Michel Brankart, and Jacques Verron.
\newblock {Wide-Swath Altimetric Satellite Data Assimilation With Correlated-Error Reduction}.
\newblock \emph{Frontiers in Marine Science}, 6:\penalty0 822, 1 2020.
\newblock ISSN 22967745.
\newblock \doi{10.3389/FMARS.2019.00822/BIBTEX}.

\bibitem[Fresnay et~al.(2018)Fresnay, Ponte, Le~Gentil, and Le~Sommer]{Fresnay2018ReconstructionAtlantic}
S.~Fresnay, A.~L. Ponte, S.~Le~Gentil, and J.~Le~Sommer.
\newblock {Reconstruction of the 3-D Dynamics From Surface Variables in a High-Resolution Simulation of North Atlantic}.
\newblock \emph{Journal of Geophysical Research: Oceans}, 123\penalty0 (3):\penalty0 1612--1630, 3 2018.
\newblock ISSN 2169-9291.
\newblock \doi{10.1002/2017JC013400}.
\newblock URL \url{https://onlinelibrary.wiley.com/doi/full/10.1002/2017JC013400 https://onlinelibrary.wiley.com/doi/abs/10.1002/2017JC013400 https://agupubs.onlinelibrary.wiley.com/doi/10.1002/2017JC013400}.

\bibitem[Amores et~al.(2018)Amores, Jord{\`{a}}, Arsouze, and Le~Sommer]{Amores2018UpProducts}
Angel Amores, Gabriel Jord{\`{a}}, Thomas Arsouze, and Julien Le~Sommer.
\newblock {Up to What Extent Can We Characterize Ocean Eddies Using Present-Day Gridded Altimetric Products?}
\newblock \emph{Journal of Geophysical Research: Oceans}, 123\penalty0 (10):\penalty0 7220--7236, 10 2018.
\newblock ISSN 2169-9291.
\newblock \doi{10.1029/2018JC014140}.
\newblock URL \url{https://onlinelibrary.wiley.com/doi/full/10.1029/2018JC014140 https://onlinelibrary.wiley.com/doi/abs/10.1029/2018JC014140 https://agupubs.onlinelibrary.wiley.com/doi/10.1029/2018JC014140}.

\bibitem[Ruiz et~al.(2009)Ruiz, Pascual, Garau, Faug{\`{e}}re, Alvarez, and Tintor{\'{e}}]{Ruiz2009MesoscaleData}
Simón Ruiz, Ananda Pascual, Bartolomé Garau, Yannice Faug{\`{e}}re, Alberto Alvarez, and Joaquín Tintor{\'{e}}.
\newblock {Mesoscale dynamics of the Balearic Front, integrating glider, ship and satellite data}.
\newblock \emph{Journal of Marine Systems}, 78\penalty0 (SUPPL. 1):\penalty0 S3--S16, 11 2009.
\newblock ISSN 0924-7963.
\newblock \doi{10.1016/J.JMARSYS.2009.01.007}.

\bibitem[Volpe et~al.(2019)Volpe, Colella, Brando, Forneris, La~Padula, Di~Cicco, Sammartino, Bracaglia, Artuso, and Santoleri]{Volpe2019MediterraneanProcessing}
Gianluca Volpe, Simone Colella, Vittorio~E. Brando, Vega Forneris, Flavio La~Padula, Annalisa Di~Cicco, Michela Sammartino, Marco Bracaglia, Florinda Artuso, and Rosalia Santoleri.
\newblock {Mediterranean ocean colour Level 3 operational multi-sensor processing}.
\newblock \emph{Ocean Science}, 15\penalty0 (1):\penalty0 127--146, 2 2019.
\newblock ISSN 18120792.
\newblock \doi{10.5194/OS-15-127-2019}.

\bibitem[Zarokanellos et~al.(2022)Zarokanellos, Rudnick, Garcia-Jove, Mourre, Ruiz, Pascual, and Tintor{\'{e}}]{Zarokanellos2022FrontalObservations}
Nikolaos~D. Zarokanellos, Daniel~L. Rudnick, Maximo Garcia-Jove, Baptiste Mourre, Simon Ruiz, Ananda Pascual, and Joaquin Tintor{\'{e}}.
\newblock {Frontal Dynamics in the Alboran Sea: 1. Coherent 3D Pathways at the Almeria-Oran Front Using Underwater Glider Observations}.
\newblock \emph{Journal of Geophysical Research: Oceans}, 127\penalty0 (3):\penalty0 e2021JC017405, 3 2022.
\newblock ISSN 2169-9291.
\newblock \doi{10.1029/2021JC017405}.
\newblock URL \url{https://onlinelibrary.wiley.com/doi/full/10.1029/2021JC017405 https://onlinelibrary.wiley.com/doi/abs/10.1029/2021JC017405 https://agupubs.onlinelibrary.wiley.com/doi/10.1029/2021JC017405}.

\bibitem[Heslop et~al.(2012)Heslop, Ruiz, Allen, L{\'{o}}pez-Jurado, Renault, and Tintor{\'{e}}]{Heslop2012AutonomousSea}
Emma~E. Heslop, Simón Ruiz, John Allen, José~Luís L{\'{o}}pez-Jurado, Lionel Renault, and Joaquín Tintor{\'{e}}.
\newblock {Autonomous underwater gliders monitoring variability at “choke points” in our ocean system: A case study in the Western Mediterranean Sea}.
\newblock \emph{Geophysical Research Letters}, 39\penalty0 (20), 10 2012.
\newblock ISSN 1944-8007.
\newblock \doi{10.1029/2012GL053717}.
\newblock URL \url{https://onlinelibrary.wiley.com/doi/full/10.1029/2012GL053717 https://onlinelibrary.wiley.com/doi/abs/10.1029/2012GL053717 https://agupubs.onlinelibrary.wiley.com/doi/10.1029/2012GL053717}.

\bibitem[Ruiz et~al.(2019)Ruiz, Claret, Pascual, Olita, Troupin, Capet, Tovar-S{\'{a}}nchez, Allen, Poulain, Tintor{\'{e}}, and Mahadevan]{Ruiz2019EffectsPhytoplankton}
Simón Ruiz, Mariona Claret, Ananda Pascual, Antonio Olita, Charles Troupin, Arthur Capet, Antonio Tovar-S{\'{a}}nchez, John Allen, Pierre~Marie Poulain, Joaquín Tintor{\'{e}}, and Amala Mahadevan.
\newblock {Effects of Oceanic Mesoscale and Submesoscale Frontal Processes on the Vertical Transport of Phytoplankton}.
\newblock \emph{Journal of Geophysical Research: Oceans}, 124\penalty0 (8):\penalty0 5999--6014, 8 2019.
\newblock ISSN 2169-9291.
\newblock \doi{10.1029/2019JC015034}.
\newblock URL \url{https://onlinelibrary.wiley.com/doi/full/10.1029/2019JC015034 https://onlinelibrary.wiley.com/doi/abs/10.1029/2019JC015034 https://agupubs.onlinelibrary.wiley.com/doi/10.1029/2019JC015034}.

\bibitem[Alvarez et~al.(2007)Alvarez, Garau, and Caiti]{Alvarez2007CombiningOcean}
Alberto Alvarez, Bartolome Garau, and Andrea Caiti.
\newblock {Combining networks of drifting profiling floats and gliders for adaptive sampling of the Ocean}.
\newblock \emph{Proceedings - IEEE International Conference on Robotics and Automation}, pages 157--162, 2007.
\newblock ISSN 10504729.
\newblock \doi{10.1109/ROBOT.2007.363780}.

\bibitem[Troupin et~al.(2015)Troupin, Beltran, Heslop, Torner, Garau, Allen, Ruiz, and Tintor{\'{e}}]{Troupin2015AManagement}
C.~Troupin, J.~P. Beltran, E.~Heslop, M.~Torner, B.~Garau, J.~Allen, S.~Ruiz, and J.~Tintor{\'{e}}.
\newblock {A toolbox for glider data processing and management}.
\newblock \emph{Methods in Oceanography}, 13-14:\penalty0 13--23, 9 2015.
\newblock ISSN 2211-1220.
\newblock \doi{10.1016/J.MIO.2016.01.001}.

\bibitem[Miyoshi and Kondo(2013)]{Miyoshi2013AFilter}
Takemasa Miyoshi and Keiichi Kondo.
\newblock {A Multi-Scale Localization Approach to an Ensemble Kalman filter}.
\newblock \emph{SOLA}, 9\penalty0 (1):\penalty0 170--173, 2013.
\newblock ISSN 1349-6476.
\newblock \doi{10.2151/SOLA.2013-038}.

\bibitem[Chen et~al.(2021)Chen, Li, Luna, Chung, Rowe, Du, Solnes, and Frey]{Chen2021LearningNetworks}
Junyu Chen, Ye~Li, Licia~P. Luna, Hyun~W. Chung, Steven~P. Rowe, Yong Du, Lilja~B. Solnes, and Eric~C. Frey.
\newblock {Learning fuzzy clustering for SPECT/CT segmentation via convolutional neural networks}.
\newblock \emph{Medical physics}, 48\penalty0 (7):\penalty0 3860--3877, 7 2021.
\newblock ISSN 2473-4209.
\newblock \doi{10.1002/MP.14903}.
\newblock URL \url{https://pubmed.ncbi.nlm.nih.gov/33905560/}.

\end{thebibliography}






\end{document}